\documentclass[runningheads]{llncs}
\usepackage[T1]{fontenc}
\usepackage{graphicx}
\graphicspath{ {images/} }

\usepackage{array}
\newcolumntype{P}[1]{>{\centering\arraybackslash}p{#1}}

\begin{document}
\title{A Humanoid Social Robot as a Teaching Assistant in the Classroom}
%
%
\author{Thomas Sievers\orcidID{0000-0002-8675-0122}}
%

\institute{Institute of Information Systems, University of Lübeck,\\ Ratzeburger Allee 160, 23562 Lübeck, Germany\\
\email{t.sievers@uni-luebeck.de}}

%
\maketitle              
\begin{abstract}
Although innovation and the support of new technologies are much needed to ease the burden on the education system, social robots in schools to help teachers with educational tasks are rare. Child-Robot Interaction (CRI) could support teachers and add an embodied social component to modern multi-modal and multi-sensory learning environments already in use. The social robot Pepper, connected to the Large Language Model (LLM) ChatGPT, was used in a high school classroom to teach new learning content to groups of students. I tested the technical possibilities with the robot on site and asked the students about their acceptance and perceived usefulness of teaching with the help of a social robot. All participants felt that the robot's presentation of the learning material was appropriate or at least partially appropriate and that its use made sense.

\end{abstract}
\section{Introduction}

Although innovation and the support of technology are urgently needed to ease the burden on the education system, especially in today's world, social robots are still rare in children's school environments. But with Child–Robot Interaction (CRI) they can be effective and helpful tools that support children's learning. They achieve better learning results than virtual agents and are comparable to human tutors if the tasks are simple and social \cite{Severson}.

Children learn more easily in multimodal and multi-sensory environments. Artificial Intelligence (AI) and Large Language Models (LLMs) can serve as tools for this purpose. However, AI systems developed for children are often based on individual activities on a screen and ignore an essential element of child development: interaction with a social environment. If children learn or solve a problem together, they retain what they have learned better and at the same time develop numerous other skills such as empathy, theory of mind, metacognition and emotion regulation \cite{ImpactEU}. There is also good experience with social robots that help children with autism spectrum disorders to learn while taking their special needs into account \cite{Yousif,Qidwai}.

Language education, robotics education, teaching assistance, social skills development and special education as well as guided learning through feedback are named as the five most important applications of educational robots, and preschools and primary schools are seen as having a large potential for the use of educational robots in the near future \cite{CHENG2018399,SieversAAAI}. Social robots are used as moderators in a collaborative learning process and can create a pleasant learning experience for learners \cite{Buchem23,Buchem24}. Children's valence and engagement in language learning can be improved through personalized tutoring from a robot, even when used in the natural environment of preschool classrooms during regular activities \cite{Gordon}.

One major challenge when using social robots as intelligent learning systems in the classroom is to combine the robot-centered perspective, i.e. what robots are technically capable of, with the point of view of the child-centered perspective, which represents how the child or children can benefit from the robot and how the robot should act in order to best support them in achieving the interaction goals \cite{Rudenko}. Even if a human who is criticized by a robot tends to have a negative reaction to the robot when it gives a poor evaluation, the evaluation by a robot instead of a human teacher could be in the robot's favor because a robot tutor is free from bias toward the students' gender, ethnicity, socioeconomic status, personal preferences, or other considerations \cite{You11,Smakman}.

In this paper, I summarize findings from testing the humanoid social robot Pepper in interaction with students at a high school in Germany. The robot's main task was to support lessons by presenting new learning content in conversation with small groups of students. To make the robot speak I applied the LLM of OpenAI's Generative Pretrained Transformer (GPT, commonly known as ChatGPT) \cite{OpenAI}. I chose the approach of on-site practical testing for the question of the feasibility of using a humanoid social robot as a teacher's assistant in a normal school classroom.

\section{Related work}

There are many studies on the suitability of robots in education for children of different age groups with various focuses. Woo et al. investigated field studies on the use of social robots in real-life classrooms, which showed that the use of social robots in natural school environments is feasible, but that there are difficulties in personalizing the interaction, among other things \cite{WOO2021100388}. 
Various types of breakdowns, including misunderstandings, malfunctions or inconsistencies in the interaction between children and a social robot tutor at an elementary school were identified by Serholt \cite{SERHOLT2018250}. Other studies suggested that social robots could be used as language tutors for children from an early age \cite{Belpaeme2018,Rianne,Konijn2022}. This seems to be one way of compensating for language deficits in the school environment.

Personalization and adaptation on the part of the tutor to the student's current abilities appear to make sense. However, Gao et al. used a Pepper robot as a tutor to help people solve logic puzzles. Their results showed that the personalization of the robot's behavior on user's task performance had a negative influence on people's perception. A more varied behavior of the robot was preferred \cite{8525832}. They came to the conclusion that caution is required when developing social and adaptive behaviors in robots designed to support human learning.

A review conducted by DiPaola et al. on social robots and children's rights found that in Human-Robot Interaction (HRI), nonverbal interaction related to the physical nature of robots is considered as important as verbal interaction and that robots are differentiated from other AI-based devices by their embodiment \cite{DiPaola}. Most of the papers examined in the review raised concerns about the physical safety of children, and only a few papers referred to the psychological or mental safety of children. While inclusion seems to be one of the most popular research topics in the field of children's rights, other aspects such as explainability and fairness have not yet been sufficiently researched.

Concerns that a robot tutor could have a negative impact on children's psychological well-being and happiness, as well as trust and privacy, may be offset by evidence of great opportunities in terms of improving educational outcomes and freedom from bias \cite{Smakman}. Apart from particular approaches for inclusion, autism, language learning for migrants or similar special cases, my interest lies in the use of a social robot in the classroom at a typical school with ordinary students.

\section{Methods}

Studies show that the acceptance of robots by children depends to a large extent on the age of the child and the characteristics of the robot \cite{Severson}. In a real school environment, it is technically very difficult to enable a robot to understand conversations with people in background noise or in groups using today's technology. This complicates the realization of real social interactions immensely \cite{LeTendre}. However, the possibility of using ChatGPT to realize the robot's communication capabilities has significantly reduced many limitations in dialog with humans, since in principle at least conversations on all conceivable topics are now possible without further implementation effort.

In order to investigate how the robot and in particular learning with this robot was perceived by the students of a ninth grade of a German high school, I asked for feedback on my third visit. For this I used the 18-item Robotic Social Attributes Scale (RoSAS) \cite{Carpinella}. It comprises three underlying scale dimensions -- the factors \textit{warmth}, \textit{competence} and \textit{discomfort}. RoSAS is based on items from the Godspeed Scale \cite{Bartneck} and psychological literature on social perception. It is considered to be a psychologically valid scale of robotic perception. The questionnaire was supplemented by a query on how the participants assess the suitability of this robot for adequately teaching the subject matter and whether they consider learning with the robot to be a useful addition to lessons.

\subsection{Humanoid Social Robot Pepper}

\begin{figure}
  \centering
  \includegraphics[scale=0.16]{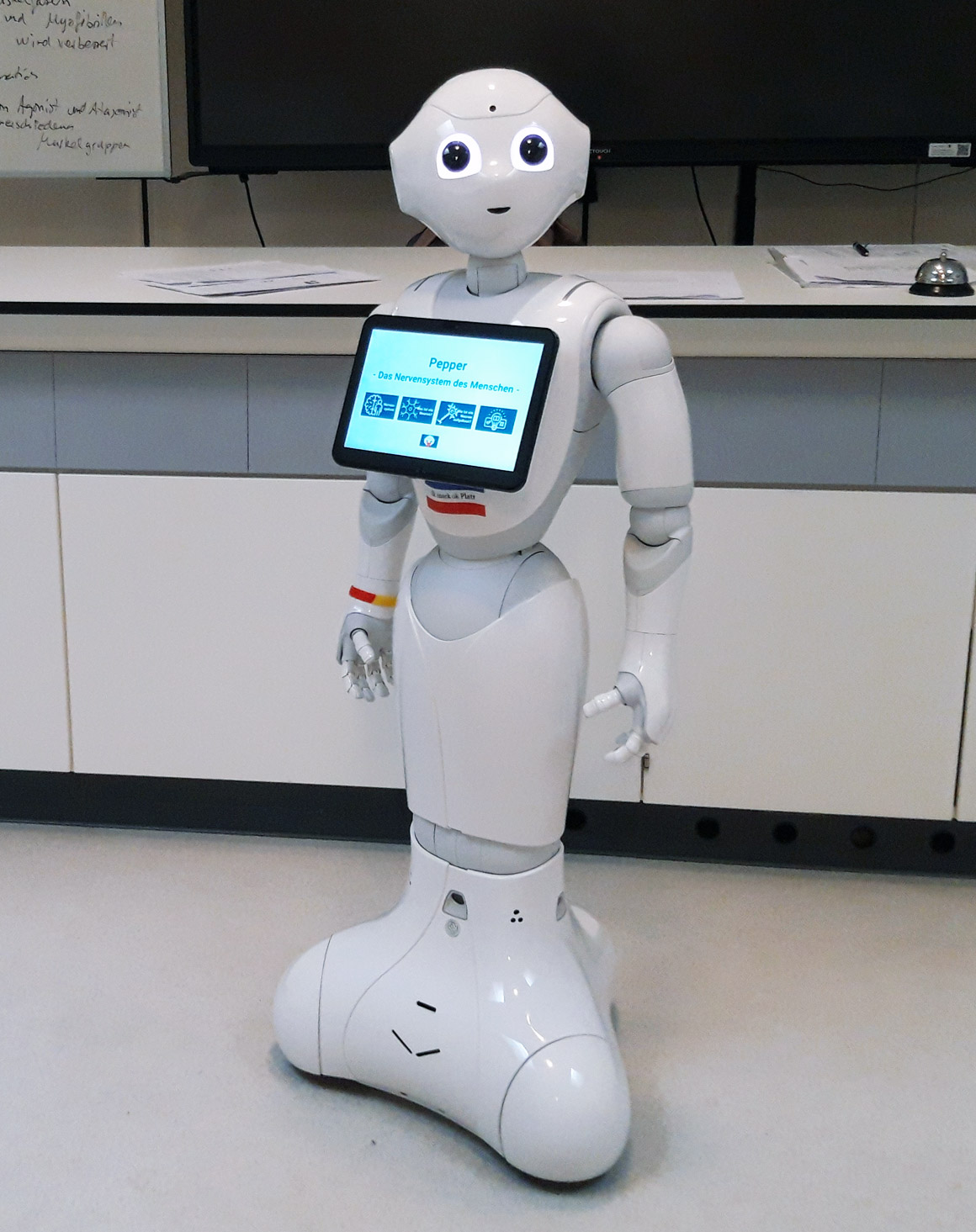}
  \caption{Robot Pepper in a classroom ready for action}
  \label{fig_moin-pepper}
\end{figure}

The humanoid social robot Pepper, shown in Figure~\ref{fig_moin-pepper}, was developed by Aldebaran and first released in 2015 \cite{Pepper}. The robot is 120 centimeters tall and optimized for interactions with humans. It is able to engage with people through conversation, gestures and its tablet. I used the tablet to select the lesson material and to display what Pepper is currently saying. The robot features an open and fully programmable platform so that developers can program their own applications to run on Pepper.

In dialogs with humans, my robot application forwarded utterances of the human dialog partner as input to the OpenAI Application Programming Interface (API), which returned an answer as response of the API. With each API call, the entire dialog was transferred to the GPT model. This allowed the model to constantly `remember' what was previously said and refer to it as the dialog progresses. In this way, complex dialogs became possible.

\subsection{Lesson Support and Assistance to the Teacher}

To test ways of supporting teachers in normal school lessons, I took the robot three times to a ninth grade biology class at a high school with about 18 students. The topic of the lessons at the time was the human nervous system. In order to integrate the robot into normal lessons in a meaningful way, it was necessary to inform the robot and the GPT model system about the material to be covered. This was done by feeding the learning material distributed to the students by the teacher to the ChatGPT system prompt, ensuring that the language model focused on the content relevant to the lesson. The use of ChatGPT offered the opportunity for students to ask questions or receive additional information beyond the material provided -- even if this material was then generally taught without supervision by a teacher.

The interactions took place in small groups in a separate room, with one group learning with the robot and the other groups in the classroom with the teacher. After about 10 minutes, the groups swapped. It was important that the interacting students were close to the robot in order to optimize communication. I did not implement any special animations or context-dependent gestures that would be possible for Pepper in principle. However, the robot constantly moved its head, upper body and arms a little and automatically supported its utterances with smaller gestures to show a basic liveliness.

One form of interaction offered the opportunity to test the newly acquired knowledge with a quiz in which the robot named facts from the learning material that could be incorrect. The students had to judge which statements were correct and which were not. As a short answer from the students was sufficient, there were hardly any complications in terms of communication.

\begin{figure}
  \centering
  \includegraphics[scale=0.16]{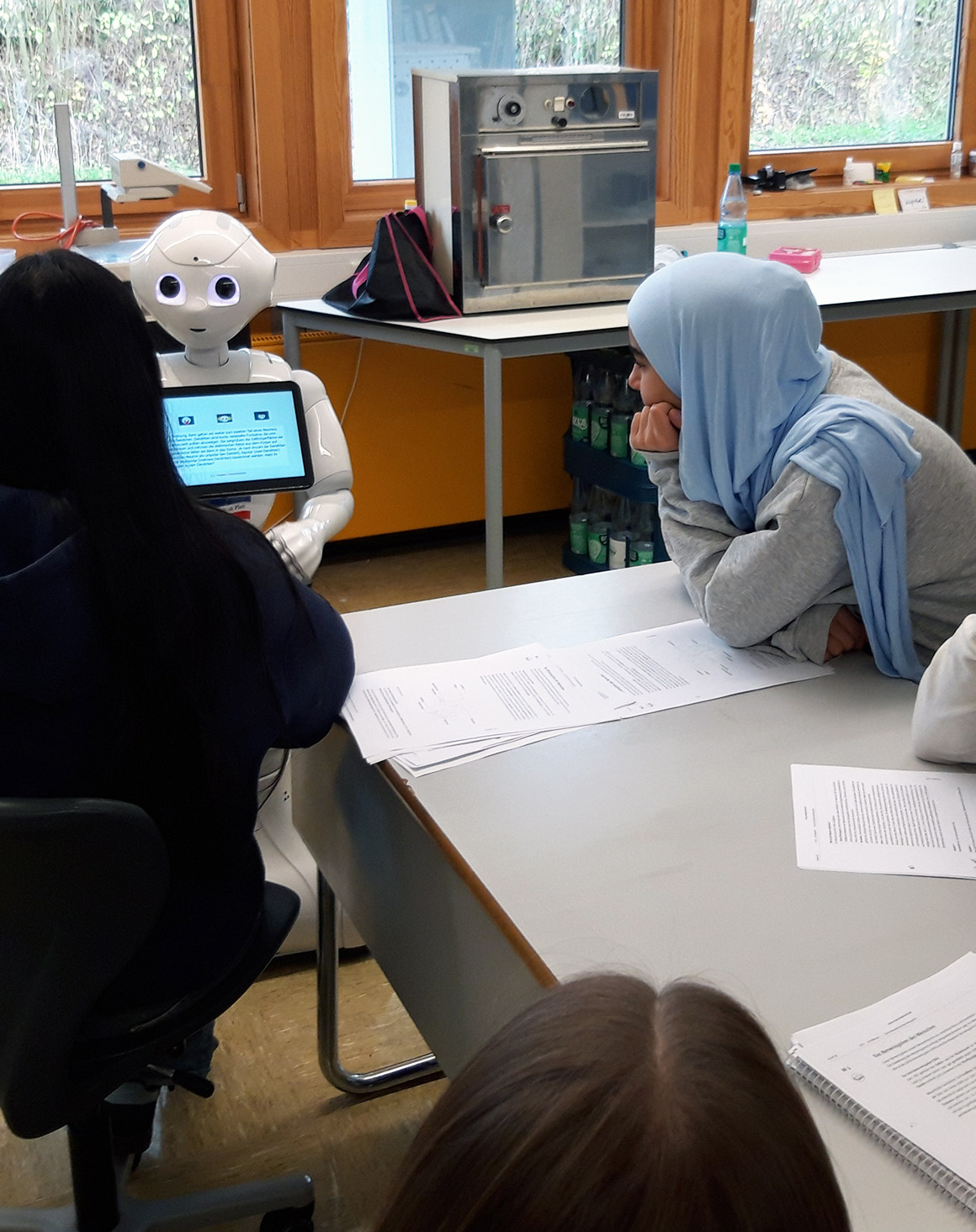}
  \caption{Robot Pepper in dialog with students to teach current learning content at a high school}
  \label{fig_trave-gym2}
\end{figure}

\subsection{Experimental Setup}

To interact and learn new content about the biology of the human nervous system, the students sat opposite the humanoid social robot Pepper and had a face-to-face conversation in a group of 4-5 people. All dialogs were conducted in German and the prompts for the GPT-4 model were also given in German.

Zero-shot prompting was used for the system role to instruct GPT-4 to execute the tasks as a completion task \cite{Brown}. The general system prompt was: \textit{`You are a robot teaching biology, specifically about the human nervous system, to high school students. You ask the class questions on this topic and refer to the facts of the following learning content, but also answer the students' questions. Never explain several things in one answer, but divide the learning content into small sections so that it remains easy to understand.'}, followed by the actual learning content in text form. The latter included information on the structure of the nervous system or the structure of the neuron, for example. The desired learning content could be started via a button on Pepper's tablet. The robot then guided the students through the material in dialog and answered questions even beyond the given text.

\section{Results and Learnings}

The fact that the robot was able to answer questions and explain issues in more detail than the material provided in the form of learning content and task sheets was generally considered to be useful for providing support in lessons. One student explicitly asked the robot to summarize and repeat the learning content again in order to take notes. Most of the students liked the robot's appearance, the liveliness and attentiveness, but some students certainly questioned the purpose of using the robot, for example, why they should not simply use their cell phone instead of asking the robot for a short research task. After all, this would be quicker. Nevertheless, they remained interested and willing to work with the robot over a number of teaching units spread over several weeks.

A major problem was the inability of the (Pepper) robot to reliably understand what the students were saying. This was often not achieved or only incompletely. It was also a general problem for the robot to distinguish between pauses in a sentence and pauses at the end of an utterance for turn-taking, which is why humans should avoid pauses e.g. for thought in their sentences when talking to the robot. This type of concentrated speaking without a pause in the sentence is not only for children difficult to master. The use of alternative speech-to-text converters would perhaps be one way of getting to grips with this problem \cite{SieversSii}. With a little practice and experience, it is possible to adapt one's own way of speaking through clearer articulation and speaking loudly without pausing in the phrase so that the robot understands the spoken sentence better. But there is still a long way to go to achieve a truly human-like receptiveness.

A quiz on the learning content, in which the robot states true or false facts, proved to be a practical way of checking learning success. The students then only had to comment on the statement as true or false. This meant that the students' statements were short and clear, making them easy for the robot to understand. Longer and more complex student statements increased the likelihood of the robot not understanding or misunderstanding.

\subsection{Feedback from a ninth grade at a high school biology class}

On a third visit to a ninth grade class at a German high school, I asked the students for feedback on their experience of being taught biology by a social robot. 9 female and 9 male participants aged between 14 and 17 took part in the robot interaction in a group of 4-5 people on this third visit. All of them had already met the robot at at least one previous visit. They were therefore already familiar with it. The robot had previously explained something about \textit{blood groups} to them in the learning session.

After each group interaction with the robot, the 18 items of the RoSAS were presented to the participants to assess the dialog and their perception of the robot. They were asked, \textit{“Using the scale provided, how closely are the words below associated with your perception of the robot?”}. The participants responded using a 5-point Likert scale from 1 = \textit{does not apply at all} to 5 = \textit{applies}. Every RoSAS factor comprises six items. For the factor \textit{warmth} they are: happy, feeling, social, organic, compassionate and emotional. \textit{Competence} includes: capable, responsive, interactive, reliable, competent and knowledgable. And \textit{discomfort} comprises: scary, strange, awkward, dangerous, awful and aggressive. Each factor consisting of six items could therefore receive a total of between 6 and 30 points, with a score of 6 representing the lowest agreement and 30 the highest. Apart from year of birth and gender, no personal data was collected. In addition to the RoSAS questionnaire, I also asked the following questions:

Question 1: \textit{`Do you have the impression that the robot was able to convey the learning material to you appropriately?'}

Question 2: \textit{`Did you find learning with the robot a useful addition to your lessons?'}

Student feedback on the two questions is shown in Tables \ref{question1} and \ref{question2}. All participants of the learning interaction were of the opinion that the presentation of the learning material by the robot was appropriate or at least partially appropriate and that its use made sense.

\begin{table}[]
\centering
\caption{Do you have the impression that the robot was able to convey the learning material to you appropriately?}
\label{question1}
\begin{tabular}{|P{1.5cm}|P{1.5cm}|P{1.5cm}|P{1.5cm}|}
\hline
       & yes & no & in part \\ \hline
female & 4   & 0  & 5       \\ \hline
male   & 5   & 0  & 4       \\ \hline
\end{tabular}
\end{table}

\begin{table}[]
\centering
\caption{Did you find learning with the robot a useful addition to your lessons?}
\label{question2}
\begin{tabular}{|P{1.5cm}|P{1.5cm}|P{1.5cm}|P{1.5cm}|}
\hline
       & yes & no & in part \\ \hline
female & 9   & 0  & 0       \\ \hline
male   & 7   & 0  & 2       \\ \hline
\end{tabular}
\end{table}

\begin{figure}
  \includegraphics[width=1.0\textwidth]{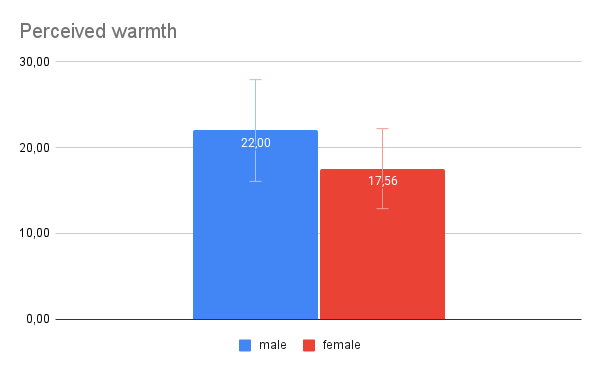}
  \caption{The results of the perceived warmth for the robot as an assistant teacher. Error bars show standard deviations.}
  \label{fig_perceived-warmth}
\end{figure}

Figure~\ref{fig_perceived-warmth} shows the results with regard to the factor \textit{warmth} according to the evaluation of the RoSAS questionnaire. The corresponding items happy, feeling, social, organic, compassionate and emotional ware rated with an average of 22.00 out of a total of 30 possible points by male participants, and with an average of 17.56 by the female students. It might be interesting to delve deeper here and try to find out why there is a noticeable difference in the perception of male and female students for this (and only this) factor.

\begin{figure}
  \includegraphics[width=1.0\textwidth]{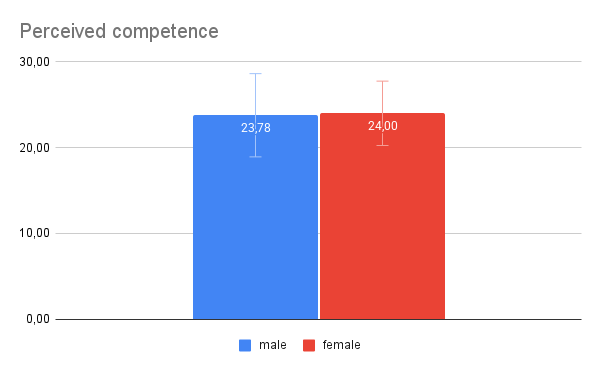}
  \caption{The results of the perceived competence for the robot as an assistant teacher. Error bars show standard deviations.}
  \label{fig_perceived-competence}
\end{figure}

\begin{figure}
  \includegraphics[width=1.0\textwidth]{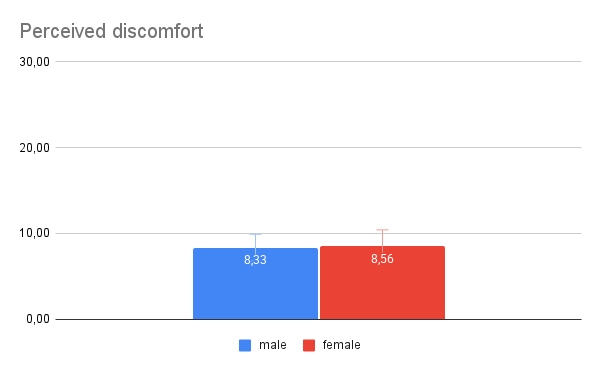}
  \caption{The results of the perceived discomfort for the robot as an assistant teacher. Error bars show standard deviations.}
  \label{fig_perceived-discomfort}
\end{figure}

Figure~\ref{fig_perceived-competence} shows the results for perceived \textit{competence} including the items capable, responsive, interactive, reliable, competent and knowledgable. These items were generally perceived as quite high by both male and female students at around 24 points. Our results in Figure~\ref{fig_perceived-discomfort} on the factor \textit{discomfort}, which includes the items scary, strange, awkward, dangerous, awful and aggressive, show rather low values of around 8-9 points in the assessment of the robot by both male and female students.

In my opinion, the relatively high sympathy ratings and the high level of competence awarded are good prerequisites for working as an assistant teacher. The competence values appear to be little or not at all affected by the frequent communication breakdowns. Presumably the perceived competence as a teacher based on specialist knowledge predominates here.
When asked if they find learning with the robot a useful addition to their lessons the students made the following statements, for example:

\begin{itemize}

\item `It made sense, as it was a change from the normal lessons.'
\item `It's something different, but there was too much information at the same time.'
\item `The robot can explain the topic slightly differently than the teacher, which is sometimes better and easier to understand.'
\item `If the system is worked out a bit more, that's a good idea.'

\end{itemize}

My impression from these results is that a social robot can be put to good use in the classroom. Suitable application scenarios would depend on the grade level or age of the children and, of course, the school subject. Some tasks are easier to implement and use, for example language training or a quiz on the content to learn with short answer options for the children, while others would be difficult to use due to the currently inadequate technology, for example conducting extensive discussions with longer verbal contributions from the students, which the robot may not be able to pick up correctly. It would also require some imagination on the part of the teachers together with robot experts to create suitable applications and concepts for use in the classroom.

\section{Conclusion}

Social robots can complement multi-modal and multi-sensory learning environments as a form of embodied and social AI. They offer unique and highly interesting facets of interaction face-to-face or in small groups for children. However, today's robotic systems are not yet robust enough in every respect for autonomous continuous use in real learning environments. But despite the difficulties and glitches in verbal communication, for example, students find such systems helpful, interesting and useful. Further technical developments could help to solve the problems in the medium term. By then at the latest, robots as assistant tutors should actually be able to relieve teachers of certain teaching tasks. It seems appropriate to address this topic in schools and society now. However, in my experience with schools and in conversations with teachers on this topic, the conclusion is that, at least in Germany, it depends almost exclusively on the individual commitment of individual teachers as to whether the subject of AI or innovative approaches such as the use of social robots takes place in lessons. An overall concept in which all schools could participate is not evident.

%
%
%
\bibliographystyle{splncs04}
\bibliography{references}

\begin{thebibliography}{10}
\providecommand{\url}[1]{\texttt{#1}}
\providecommand{\urlprefix}{URL }
\providecommand{\doi}[1]{https://doi.org/#1}

\bibitem{Pepper}
{Aldebaran, United Robotics Group and Softbank Robotics}: Pepper. Tech. rep.
  (2025), \url{https://www.aldebaran.com/en/pepper}

\bibitem{Bartneck}
Bartneck, C., Kulic, D., Croft, E., Zoghbi, S.: Measurement instruments for the
  anthropomorphism, animacy, likeability, perceived intelligence, and perceived
  safety of robots. International Journal of Social Robotics  \textbf{1},
  71--81 (01 2008). \doi{10.1007/s12369-008-0001-3}

\bibitem{Belpaeme2018}
Belpaeme, T., Vogt, P., van~den Berghe, R., Bergmann, K., Göksun, T., de~Haas,
  M., Kanero, J., Kennedy, J., Küntay, A.C., Oudgenoeg-Paz, O., Papadopoulos,
  F., Schodde, T., Verhagen, J., Wallbridge, C.D., Willemsen, B., de~Wit, J.,
  Geçkin, V., Hoffmann, L., Kopp, S., Krahmer, E., Mamus, E., Montanier, J.M.,
  Oranç, C., Pandey, A.K.: Guidelines for designing social robots as second
  language tutors. International Journal of Social Robotics  \textbf{10},
  325--341 (06 2018). \doi{10.1007/s12369-018-0467-6}

\bibitem{Rianne}
van~den Berghe, R., Verhagen, J., Oudgenoeg-Paz, O., van~der Ven, S., Leseman,
  P.: Social robots for language learning: A review. Review of Educational
  Research  \textbf{89}(2),  259--295 (2019). \doi{10.3102/0034654318821286}

\bibitem{Brown}
Brown, T., Mann, B., Ryder, N., Subbiah, M., Kaplan, J., Dhariwal, P.,
  Neelakantan, A., Shyam, P., Sastry, G., Askell, A., Agarwal, S.,
  Herbert-Voss, A., Krueger, G., Henighan, T., Child, R., Ramesh, A., Ziegler,
  D., Wu, J., Winter, C., Amodei, D.: Language models are few-shot learners (05
  2020). \doi{10.48550/arXiv.2005.14165}

\bibitem{Buchem23}
Buchem, I.: Scaling-up social learning in small groups with robot supported
  collaborative learning (rscl): Effects of learners’ prior experience in the
  case study of planning poker with the robot nao. Applied Sciences
  \textbf{13}, ~4106 (03 2023). \doi{10.3390/app13074106}

\bibitem{Buchem24}
Buchem, I., Sostak, S., Christiansen, L.: Human–robot co-facilitation in
  collaborative learning: A comparative study of the effects of human and robot
  facilitation on learning experience and learning outcomes. J —
  Multidisciplinary Scientific Journal  \textbf{7},  236--263 (07 2024).
  \doi{10.3390/j7030014}

\bibitem{Carpinella}
Carpinella, C.M., Wyman, A.B., Perez, M.A., Stroessner, S.J.: The robotic
  social attributes scale (rosas): Development and validation. In: 2017 12th
  ACM/IEEE International Conference on Human-Robot Interaction (HRI. pp.
  254--262 (2017). \doi{10.1145/2909824.3020208}

\bibitem{CHENG2018399}
Cheng, Y.W., Sun, P.C., Chen, N.S.: The essential applications of educational
  robot: Requirement analysis from the perspectives of experts, researchers and
  instructors. Computers \& Education  \textbf{126},  399--416 (2018).
  \doi{https://doi.org/10.1016/j.compedu.2018.07.020},
  \url{https://www.sciencedirect.com/science/article/pii/S0360131518302033}

\bibitem{DiPaola}
DiPaola, D., Charisi, V., Breazeal, C., Sabanovic, S.: Children's fundamental
  rights in human-robot interaction research: A systematic review. In:
  Companion of the 2023 ACM/IEEE International Conference on Human-Robot
  Interaction. p. 561–566. HRI '23, Association for Computing Machinery, New
  York, NY, USA (2023). \doi{10.1145/3568294.3580148}

\bibitem{8525832}
Gao, Y., Barendregt, W., Obaid, M., Castellano, G.: When robot personalisation
  does not help: Insights from a robot-supported learning study. In: 2018 27th
  IEEE International Symposium on Robot and Human Interactive Communication
  (RO-MAN). pp. 705--712 (2018). \doi{10.1109/ROMAN.2018.8525832}

\bibitem{Gordon}
Gordon, G., Spaulding, S., Westlund, J.K., Lee, J.J., Plummer, L., Martinez,
  M., Das, M., Breazeal, C.: Affective personalization of a social robot tutor
  for children's second language skills. In: Proceedings of the Thirtieth AAAI
  Conference on Artificial Intelligence. p. 3951–3957. AAAI'16, AAAI Press
  (2016), \url{https://ojs.aaai.org/index.php/AAAI/article/view/9914}

\bibitem{Konijn2022}
Konijn, E.A., Jansen, B., Mondaca~Bustos, V., Hobbelink, V.L.N.F.,
  Preciado~Vanegas, D.: Social robots for (second) language learning in
  (migrant) primary school children. International Journal of Social Robotics
  \textbf{14},  827--843 (04 2022). \doi{10.1007/s12369-021-00824-3}

\bibitem{LeTendre}
LeTendre, G.K.: What social robots can teach america’s students  (2024),
  \url{https://theconversation.com/what-social-robots-can-teach-americas-students-220124}

\bibitem{OpenAI}
OpenAI: The most powerful platform for building ai productsi. Tech. rep.
  (2025), \url{https://openai.com/api/}

\bibitem{Qidwai}
Qidwai, U., Kashem, S.B.A., Conor, O.: Humanoid robot as a teacher’s
  assistant: Helping children with autism to learn social and academic skills.
  Journal of Intelligent \& Robotic Systems  \textbf{98},  759--770 (06 2020).
  \doi{10.1007/s10846-019-01075-1}

\bibitem{Rudenko}
Rudenko, I., Rudenko, A., Lilienthal, A.J., Arras, K.O., Bruno, B.: The Child
  Factor in Child–Robot Interaction: Discovering the Impact of Developmental
  Stage and Individual Characteristics, vol.~16, pp. 1879--1900 (2024).
  \doi{10.1007/s12369-024-01121-5}

\bibitem{SERHOLT2018250}
Serholt, S.: Breakdowns in children's interactions with a robotic tutor: A
  longitudinal study. Computers in Human Behavior  \textbf{81},  250--264
  (2018). \doi{https://doi.org/10.1016/j.chb.2017.12.030},
  \url{https://www.sciencedirect.com/science/article/pii/S0747563217307100}

\bibitem{Severson}
Severson, R., Peter, J., Kanda, T., Kaufman, J., Scassellati, B.: Social Robots
  and Children’s Development: Promises and Implications, pp. 627--633 (12
  2024). \doi{10.1007/978-3-031-69362-5_85}

\bibitem{SieversSii}
Sievers, T.: Silence is golden - making pauses in human utterances
  comprehensible for social robots in human-robot interaction. In: 2025
  IEEE/SICE International Symposium on System Integration (SII). pp. 526--531
  (2025). \doi{10.1109/SII59315.2025.10871032}

\bibitem{SieversAAAI}
Sievers, T.: A practical approach to child-robot interaction in the classroom.
  In: Child-AI Interaction in the Era of Foundation Models. 2025 AAAI Spring
  Symposium Series (2025 forthcoming)

\bibitem{Smakman}
Smakman, M., Konijn, E.A.: Robot tutors: Welcome or ethically questionable? In:
  Merdan, M., Lepuschitz, W., Koppensteiner, G., Balogh, R.,
  Obdr{\v{z}}{\'a}lek, D. (eds.) Robotics in Education. pp. 376--386. Springer
  International Publishing, Cham (2020)

\bibitem{ImpactEU}
{Vicky Charisi, Joint Research Centre, European Commission}: The impact of
  social robots on children’s development: What science says. European School
  Education Platform  (2022),
  \url{https://school-education.ec.europa.eu/en/discover/news/impact-social-robots}

\bibitem{WOO2021100388}
Woo, H., LeTendre, G.K., Pham-Shouse, T., Xiong, Y.: The use of social robots
  in classrooms: A review of field-based studies. Educational Research Review
  \textbf{33},  100388 (2021).
  \doi{https://doi.org/10.1016/j.edurev.2021.100388},
  \url{https://www.sciencedirect.com/science/article/pii/S1747938X21000117}

\bibitem{You11}
You, S., Nie, J., Suh, K., Sundar, S.S.: When the robot criticizes you...:
  self-serving bias in human-robot interaction. In: Proceedings of the 6th
  International Conference on Human-Robot Interaction. p. 295–296. HRI '11,
  Association for Computing Machinery, New York, NY, USA (2011).
  \doi{10.1145/1957656.1957778}

\bibitem{Yousif}
Yousif, J., Yousif, M.: Humanoid robot as assistant tutor for autistic
  children. International Journal of Computation and Applied Sciences
  \textbf{8},  8--13 (04 2020), \url{https://ssrn.com/abstract=3616810}

\end{thebibliography}

\end{document}